\documentclass[aps, twocolumn, pra, superscriptaddress, 10pt]{revtex4-2}

\usepackage{amsmath}
\usepackage{amsmath, amssymb}
\usepackage{braket}
\usepackage{todonotes}
\usepackage{ulem}
\usepackage{braket}
\usepackage{color}
\usepackage{orcidlink}
\usepackage{here}

\begin{document}

\title{Inverse design of dual-band valley-Hall topological photonic crystals with arbitrary pseudospin states}

\author{Yuki Sato}
\email[]{yuki-sato@mosk.tytlabs.co.jp}
\affiliation{Toyota Central R\&D Labs., Inc., 41-1, Yokomichi, Nagakute, Aichi, 480-1192, Japan}

\author{Shrinathan Esaki Muthu Pandara Kone}
\affiliation{Toyota Central R\&D Labs., Inc., 41-1, Yokomichi, Nagakute, Aichi, 480-1192, Japan}

\author{Junpei Oba} 
\affiliation{Toyota Central R\&D Labs., Inc., 41-1, Yokomichi, Nagakute, Aichi, 480-1192, Japan}

\author{Kenichi Yatsugi} 
\email[]{yatsugi@mosk.tytlabs.co.jp}
\affiliation{Toyota Central R\&D Labs., Inc., 41-1, Yokomichi, Nagakute, Aichi, 480-1192, Japan}

\begin{abstract}
Valley photonic crystals (VPCs) offer topological kink states that ensure robust, unidirectional, and backscattering-immune light propagation.
The design of VPCs is typically based on analogies with condensed-matter topological insulators that exhibit the quantum valley Hall effect; trial-and-error approaches are often used to tailor the photonic band structures and their topological properties, which are characterized by the local Berry curvatures.
In this paper, we present an inverse design framework based on frequency-domain analysis for VPCs with arbitrary pseudospin states. 
Specifically, we utilize the transverse spin angular momentum (TSAM) at the band edge to formulate the objective function for engineering the desired topological properties. 
Numerical experiments demonstrate that our proposed design approach can successfully produce photonic crystal waveguides exhibiting dual-band operation, enabling frequency-dependent light routing.
Our pseudospin-engineering method thus provides a cost-effective alternative for designing topological photonic waveguides, offering novel functionalities.
\end{abstract}

\maketitle

\section{Introduction} 
\vspace*{-3pt}

Photonic topological insulators (PTIs) have attracted increasing interest because of their potential to control light propagation in a topologically protected manner~\cite{ozawa2019topological, lu2014topological, khanikaev2017two,rider2019perspective}.
The topologically protected properties emerge along the interface between two topologically distinct materials and are robust against scattering even in the presence of defects or disorder.
Several types of PTIs have already been demonstrated, including those based on the quantum Hall effect (QHE), in which time-reversal symmetry is broken~\cite{raghu2008analogs, wang2009observation}; the quantum spin Hall effect (QSHE), which leverages pseudospin degrees of freedom~\cite{wu2015scheme, yang2018visualization}; and the quantum valley Hall effect (QVHE)~\cite{noh2018observation,gao2018topologically,xue2021topological}, in which the valley degree of freedom generates valley topological phases. 
Among these PTIs, valley photonic crystals (VPCs) are of particular interest because they are easily realized~\cite{iwamoto2021recent}.

In VPCs, valley kink states emerge at the interfaces between VPCs with local Berry curvatures of opposite sign near K and K’ points in momentum space~\cite{xue2021topological}. 
These kink states have been utilized in various waveguide designs, including splitters~\cite{ma2019topological}, resonators~\cite{noh2020experimental}, dual-band waveguides~\cite{chen2019valley,yatsugi2024square}, and frequency-dependent routers~\cite{tang2020frequency, wei2021frequency}.
In VPCs, pseudospin states appear as an analogue of the valley Hall state. 
The real-space phase windings (pseudospin states) are observed at the band edge at K and K’ points. 
The direction of the phase winding reflects the sign of local Berry curvature. 
This pseudospin nature is utilized for unidirectional excitation and propagation in VPCs-based waveguides~\cite{yatsugi2024square,yang2018topological,wei2021frequency}.
Furthermore, it has been revealed recently that the eigenmodes of VPCs at K and K’ points near the band gap have the meron and anti-meron spin textures~\cite{guo2020meron}.
At the core of each meron, the eigenstates display right- or left-circular polarization, directly reflecting the sign of the local Berry curvature at the corresponding K and K’ points in the Brillouin zone.
Such circularly polarized light carries non-zero transverse spin angular momentum (TSAM), which is associated with pseudospin states~\cite{deng2017transverse, shi2023dynamical}.

Conventionally,  PTIs are usually designed by analogy with electronic topological insulators (TIs) exhibiting QHE, QSHE, or QVHE.
By symmetry breaking, topological properties can be induced in photonic systems.
However, trial-and-error tailoring of photonic band structures remains necessary to achieve topologically nontrivial states.
Inverse design, particularly structural topology optimization~\cite{bendsoe1988generating, bendsoe2013topology}, addresses this challenge by systematically searching for material distributions that satisfy targeted physical properties. 
Recent studies have applied such methods to photonic and phononic TIs~\cite{christiansen2019topological, christiansen2019designing, du2020optimal, luo2021moving, chen2022inverse}.
In Refs.~\cite{christiansen2019topological, christiansen2019designing}, topological properties emerge spontaneously without any explicit requirements regarding the underlying mechanisms.
Such approaches show promise for the design of high-performance devices; however, they do not necessarily derive TIs.
In contrast, in Refs.~\cite{du2020optimal, luo2021moving}, eigenvalue-based objectives based on these mechanisms are incorporated explicitly.
Such approaches can be computationally expensive because of the need for eigenvalue analysis over the reciprocal space.

In this study, we propose inverse designs for VPCs, focusing on the TSAM around the K point.
This approach facilitates the spontaneous generation of arbitrary pseudospin states and enables the design of Berry curvatures with predetermined signs.
By focusing only on the eigenstates around the K point, our method significantly reduces the computational cost of the optimization process. 
We demonstrate the effectiveness of our approach by designing a dual-band waveguide router that enables light to propagate along frequency-dependent paths.
To the best of our knowledge, this is the first study that proposes inverse designs of VPCs with multiple band gaps and tunable topological properties; it should help expand the application of VPCs in robust waveguide devices.

The remainder of this paper is organized as follows:
In Section~\ref{sec:method}, we formulate the inverse design problem for VPCs.
Specifically, we formulate the optimization problem in the frequency domain; this spontaneously produces the eigenstate properties of VPCs.
In Section~\ref{sec:result}, we showcase two optimized photonic crystals (PCs) with two band gaps; this design enables the construction of frequency-dependent routers.
We assess the designed PCs in terms of their band structures, Berry curvatures, pseudospins, and valley kink states.
As a demonstration, we also design a dual-band waveguide that enables light to propagate along frequency-dependent paths.
The conclusions are presented in Section~\ref{sec:conclusions}.

\vspace*{-13pt}

\section{Inverse design problem} \label{sec:method}
\begin{figure}[t]
    \centering
    \includegraphics[width=0.49\textwidth]{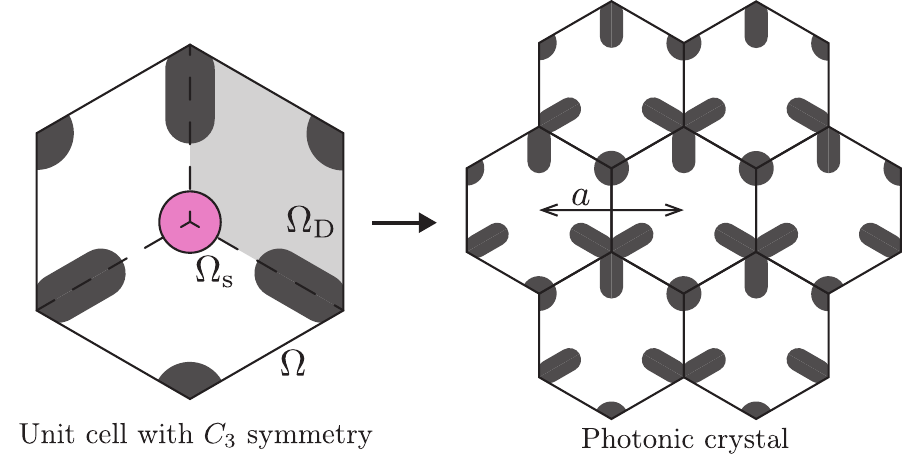}
    \caption{Conceptual diagram of the design domain (light gray) of a PC with an example material distribution (dark gray). 
    The design domain $\Omega_\mathrm{D}$ forms a unit cell $\Omega$ of a photonic crystal. 
    The unit cell has $C_3$ symmetry and lattice constant $a$. The pink domain $\Omega_\mathrm{s}$ is used to evaluate the pseudospin of the PC.}
    \label{fig:design_domain}
\end{figure}

\vspace*{-4pt}
Here, we formulate the inverse design problem for VPCs with two distinct band gaps near the target frequencies.
Figure~\ref{fig:design_domain} illustrates the design domain $\Omega_\mathrm{D}$, which constitutes the unit cell $\Omega$ of a PC with the lattice constant $a$.
We impose $C_3$ symmetry on the unit cell to create a Dirac cone at the K point, which is essential for the valley Hall effect. 
We consider the transverse magnetic (TM) -mode electromagnetic field in the frequency domain; it is governed by the Helmholtz equation for the out-of-plane electric field $E_z$:
\begin{align}
    \nabla_x^2 E_z(\boldsymbol{x}; \omega, \boldsymbol{k}) + \frac{\omega^2}{c^2} \varepsilon_\mathrm{r} E_z(\boldsymbol{x}; \omega, \boldsymbol{k}) = i \omega J_z(\boldsymbol{x}; \boldsymbol{k}),
\end{align}
where $x$ denotes the spatial coordinate, $c$ the speed of light, $\omega$ the angular frequency, and $J_z$ the external current in out-of-plane direction.
We employ the Floquet periodic boundary condition with the wave vector $\boldsymbol{k}$, i.e., $E_z(\boldsymbol{x}; \omega, \boldsymbol{k}) = E_z(\boldsymbol{x}+\boldsymbol{a}, \omega, \boldsymbol{k}) \exp(i \boldsymbol{k} \cdot \boldsymbol{a})$ for the boundary condition of the unit cell with the lattice vector $\boldsymbol{a}$.
Here, we use the notation $E_z(\boldsymbol{x}; \omega, \boldsymbol{k})$ to emphasize the dependence of the electric field on both the frequency and wave vector.
We set the external current to $J_z(\boldsymbol{x}; \boldsymbol{k}) = \exp (i \boldsymbol{k} \cdot \boldsymbol{x})$ to match the Floquet periodic boundary condition.
In the governing equation, the photonic crystal structure is represented by a spatially varying dielectric constant that depends on the fictitious material density $\gamma(\boldsymbol{x}) \in [0, 1]$.
The dielectric constant is defined on $\Omega_\mathrm{D}$ as follows:
\begin{align}
    \varepsilon_\mathrm{r}(\boldsymbol{x}; \gamma) = \varepsilon_\mathrm{r}^\mathrm{Si} \gamma(\boldsymbol{x}) + \varepsilon_\mathrm{r}^\mathrm{air} (1 - \gamma(\boldsymbol{x})),
\end{align}
where $\varepsilon_\mathrm{r}^\mathrm{Si}=12$ is the permittivity of silocon and $\varepsilon_\mathrm{r}^\mathrm{air}=1$ is the permittivity of air.
In other words, a point $\boldsymbol{x}$ is occupied by the material when $\gamma(\boldsymbol{x}) = 1$ but is void when $\gamma(\boldsymbol{x})=0$.
When $0 < \gamma(\boldsymbol{x}) < 1$, the material is regarded as an artificial intermediate material that is eliminated during density-based topology optimization using Heaviside projection~\cite{wang2011projection}.

Next, we consider the design requirements for VPCs with two band gaps around the target frequencies.
These are four requirements:
(1) Open Dirac cones exist at the K point. 
(2) Two band gaps exist between the target frequencies: the first is between $\omega_1$ and $\omega_2$ and the second is between $\omega_3$ and $\omega_4$. (3) Eigenstates at the K point have valley pseudospins.
(4) The pairs of eigenstates forming the band gaps are nearly antisymmetric.
Although these requirements involve the eigenfrequencies and their corresponding eigenstates, we evaluate them during structural optimization in the frequency-domain analysis only near the K point to avoid computationally expensive eigenvalue analysis over the first Brillouin zone.
To this end, we first define the inner product of the two electric fields as
\begin{align}
    &\braket{E_z(\boldsymbol{x}; \omega, \boldsymbol{k}), E_z(\boldsymbol{x}; \omega', \boldsymbol{k})} \nonumber \\
    &:= \int_\Omega E_z^\ast(\boldsymbol{x}; \omega, \boldsymbol{k}) E_z(\boldsymbol{x}; \omega', \boldsymbol{k}) \mathrm{d} \Omega,
\end{align}
where $\ast$ denotes the complex conjugate.
We also use the notation $\| E_z(\boldsymbol{x}; \omega, \boldsymbol{k}) \| := \sqrt{ \braket{E_z(\boldsymbol{x}; \omega, \boldsymbol{k}), E_z(\boldsymbol{x}; \omega, \boldsymbol{k})} }$.

To satisfy the first design requirement (i.e., open Dirac cone), the PC structure must have band gaps between $\omega_1$ and $\omega_2$ and between $\omega_3$ and $\omega_4$ at the K point.
This is achieved by maximizing the electric field intensities $\| E_z(\boldsymbol{x}; \omega_\mu, \boldsymbol{k}_\mathrm{K}) \|^2$ for $\mu \in \{1, 2, 3, 4\}$, where $\boldsymbol{k}_\mathrm{K}$ is the wave vector at the K point; these are maximized when $\omega_\mu$ corresponds to the eigenfrequency.
Additionally, the dispersion relation near the K point for the formation of the Dirac cones is the condition that the PC structure has the eigenfrequencies $\omega_1^{-} := \omega_1 -\Delta \omega_1$, $\omega_2^{+} := \omega_2 + \Delta \omega_2$, $\omega_3^{-} := \omega_3 - \Delta \omega_3$, and $\omega_4^{+} := \omega_4 + \Delta \omega_4$ at those points whose wave vectors are $\boldsymbol{k}_{\mathrm{K}, \Gamma} := (1-\delta)\boldsymbol{k}_\mathrm{K} + \delta \boldsymbol{k}_\Gamma$ or $\boldsymbol{k}_{\mathrm{K}, \mathrm{M}} := (1-\delta)\boldsymbol{k}_\mathrm{K} + \delta \boldsymbol{k}_\mathrm{M}$; here $\boldsymbol{k}_\Gamma$ and $\boldsymbol{k}_\mathrm{M}$ are the wave vectors at the $\Gamma$ and $M$ points, respectively.
For this study, we set $\delta=0.05$, $\omega_1=0.275~[c/2\pi a]$, $\omega_2=0.325~[c/2\pi a]$, $\omega_3=0.575~[c/2\pi a]$, $\omega_4=0.625~[c/2\pi a]$, and $\Delta \omega_1 = \Delta \omega_2 = \Delta \omega_3 = \Delta \omega_4 = 0.025~[c/2\pi a]$.
In this study, the second design requirement (i.e., formation of band gaps) was not explicitly considered because of the high computational cost.
However, it is expected to be satisfied naturally along with the first requirement (i.e., open Dirac cone).
For the third requirement (i.e, pseudospins), we evaluate the normalized TSAM at the K point as
\begin{align}
    S(\omega) := -\frac{2}{\lvert \Omega_\mathrm{s} \rvert} \mathrm{Im} \left( \int_{\Omega_\mathrm{s}} \frac{H_x^\ast(\boldsymbol{x}; \omega, \boldsymbol{k}_\mathrm{K}) H_y(\boldsymbol{x}; \omega, \boldsymbol{k}_\mathrm{K})}{ H_x^\ast H_x + H_y^\ast H_y} \mathrm{d} \Omega \right), \label{eq:polarization}
\end{align}
where $\Omega_\mathrm{s} \subset \Omega$ is the evaluation domain (see the left panel of Figure~\ref{fig:design_domain}), set as a circle with a radius of $0.1a$, and $-i \omega (H_x, H_y) = (\partial E_z / \partial y, -\partial E_z / \partial x)$, i.e., $H:=(H_x, H_y)$ is the magnetic field.
Because the normalized TSAM defined in Equation~\ref{eq:polarization} is a continuous function, it is suitable for an inverse design by density-based topology optimization, where a continuous optimization problem is solved.
Thus, we use TSAM for the third requirement (See Supplementary Information for the relationship between TSAM and pseudospin.)
The value of $S(\omega, \boldsymbol{k})$ ranges from $-1$ to $1$, and when $S(\omega, \boldsymbol{k}) = \pm 1$, the magnetic field normalized in magnitude exhibits the maximum TSAM within the domain $\Omega_\mathrm{s}$.
The positive and negative signs of $S(\omega, \boldsymbol{k})$ correspond implicitly to pseudospin-down and pseudospin-up, respectively.
While the pseudospin for the topological states is defined by the eigenstates, a spontaneous pseudospin arising in the frequency domain in response to the zero-spin input $J_z = \exp (i \boldsymbol{k} \cdot x)$ is expected to induce pseudospin in the eigenstates.
For the final design requirement (i.e., spatial inversion symmetry), the fidelity between the pairs of electric fields at the K point with one field spatially inverted is formulated as
\begin{align}
    F(\omega, \omega') := \frac{\lvert \braket{E_z(P(x); \omega, \boldsymbol{k}_\mathrm{K}), E_z(\boldsymbol{x}; \omega', \boldsymbol{k}_\mathrm{K}) }\rvert^2}{\| E_z(\boldsymbol{x}; \omega, \boldsymbol{k}_\mathrm{K}) \|^2 \| E_z(\boldsymbol{x}; \omega', \boldsymbol{k}_\mathrm{K}) \|^2},
\end{align}
where $P$ is the parity inversion operator.
When $F(\omega, \omega') = 1$, the electric fields $E_z(\boldsymbol{x}; \omega, \boldsymbol{k}_\mathrm{K})$ and $E_z(\boldsymbol{x}; \omega', \boldsymbol{k}_\mathrm{K})$ are antisymmetric, which is expected to induce antisymmetry in the pairs of eigenstates.

The optimization problem is formulated as follows:
\begin{alignat}{2}
    &\max_{\gamma} && \frac{|I|}{\sum_{(\omega, \boldsymbol{k}) \in I} \| E_z(\omega, \boldsymbol{k}) \|^{-2p}} \nonumber \\
    &&& + \frac{|I'|}{\sum_{(\omega', \boldsymbol{k}') \in I'} \| E_z(\omega', \boldsymbol{k}') \|^{-2p}} \\
    &\text{subject to:} && \nonumber \\
    &&& \lvert S(\omega_\mu) - s_\mu \rvert \leq \Delta S ~ \text{for} ~ \mu \in \{1, 2, 3, 4 \} \\
    &&& F(\omega_\mu, \omega_\nu) \geq \underline{F} ~ \text{for} ~ (\mu, \nu) \in \{(1, 2), (3, 4) \},
\end{alignat}
where $p$ is a penalty parameter, $s_\mu$ for $\mu \in \{1, 2, 3, 4 \}$ is the target pseudospin, $\Delta S$ is the allowable tolerance of the pseudospin, $\underline{F}$ is the allowable lower bound for the fidelity.
The parameters used in this study were $p=2$, $\Delta S = 0.1$, and $\underline{F} = 0.8$.
The penalty parameter $p$ places greater weight on the terms $\| E_z(\boldsymbol{x}; \omega, \boldsymbol{k}) \|^2$ with smaller values; thus, all terms can be maximized, encouraging the PC structures to be resonant for $(\omega, \boldsymbol{k}) \in I$ and $(\omega', \boldsymbol{k}') \in I'$.
The sets of frequencies and wave numbers, $I$ and $I'$, are given as follows:
\begin{align}
    I := &\left\{(\omega_1, \boldsymbol{k}_\mathrm{K}), (\omega_2, \boldsymbol{k}_\mathrm{K}), (\omega_3, \boldsymbol{k}_\mathrm{K}), (\omega_4, \boldsymbol{k}_\mathrm{K}) \right\}. \\
    I' := &\left\{(\omega_1^{-}, \boldsymbol{k}_{\mathrm{K},\Gamma}), (\omega_2^{+}, \boldsymbol{k}_{\mathrm{K}, \Gamma}), (\omega_3^{-}, \boldsymbol{k}_{\mathrm{K}, \Gamma}), (\omega_4^{+}, \boldsymbol{k}_{\mathrm{K}, \Gamma}), \right. \nonumber \\
    &\left. (\omega_1^{-}, \boldsymbol{k}_{\mathrm{K},\mathrm{M}}), (\omega_2^{+}, \boldsymbol{k}_{\mathrm{K}, \mathrm{M}}), (\omega_3^{-}, \boldsymbol{k}_{\mathrm{K}, \mathrm{M}}), (\omega_4^{+}, \boldsymbol{k}_{\mathrm{K}, \mathrm{M}}) \right\}.
\end{align}
In our study, artificial damping is set to the material by adding an imaginary part to the dielectric constant to avoid ill-posing of this optimization problem. 
Unless this damping, the values of this objective function can be infinitely large when the design is resonant.
This artificial damping also serves to prevent the gradient of the objective function from being extremely small when given frequencies are far from the eigenfrequencies of the design.
This is because the damping factor makes the Q-factor small, i.e., the electric field intensity $\| E_z(\boldsymbol{x}; \omega, \boldsymbol{k}) \|^2$ exhibits blunt peaks at the eigenfrequencies.
Specifically, we set $\mathrm{Im} \left(\varepsilon^\mathrm{Si} \right) = 1/(2Q)$ with $Q=10$ in this study.
This (heuristic) artificial damping is similar to the frequency-averaged function proposed in Ref.~\cite{liang2013formulation}, which could be evaluated using complex material constants and made the optimization problem in that study well-posed.
To ensure smoothness of the optimized PCs, partial differential equation-based filtering and Heaviside projection techniques~\cite{kawamoto2011heaviside} were employed.
Minimum length scale control using geometric constraints~\cite{zhou2015minimum} was also employed for both the solid and void regions to ensure manufacturability.
The optimization was performed using COMSOL Multiphysics software~\cite{comsol}; the globally convergent version of the method of moving asymptotes~\cite{svanberg1987method} was used as an optimizer.

\begin{figure}[t]
    \centering
    \includegraphics[width=\columnwidth]{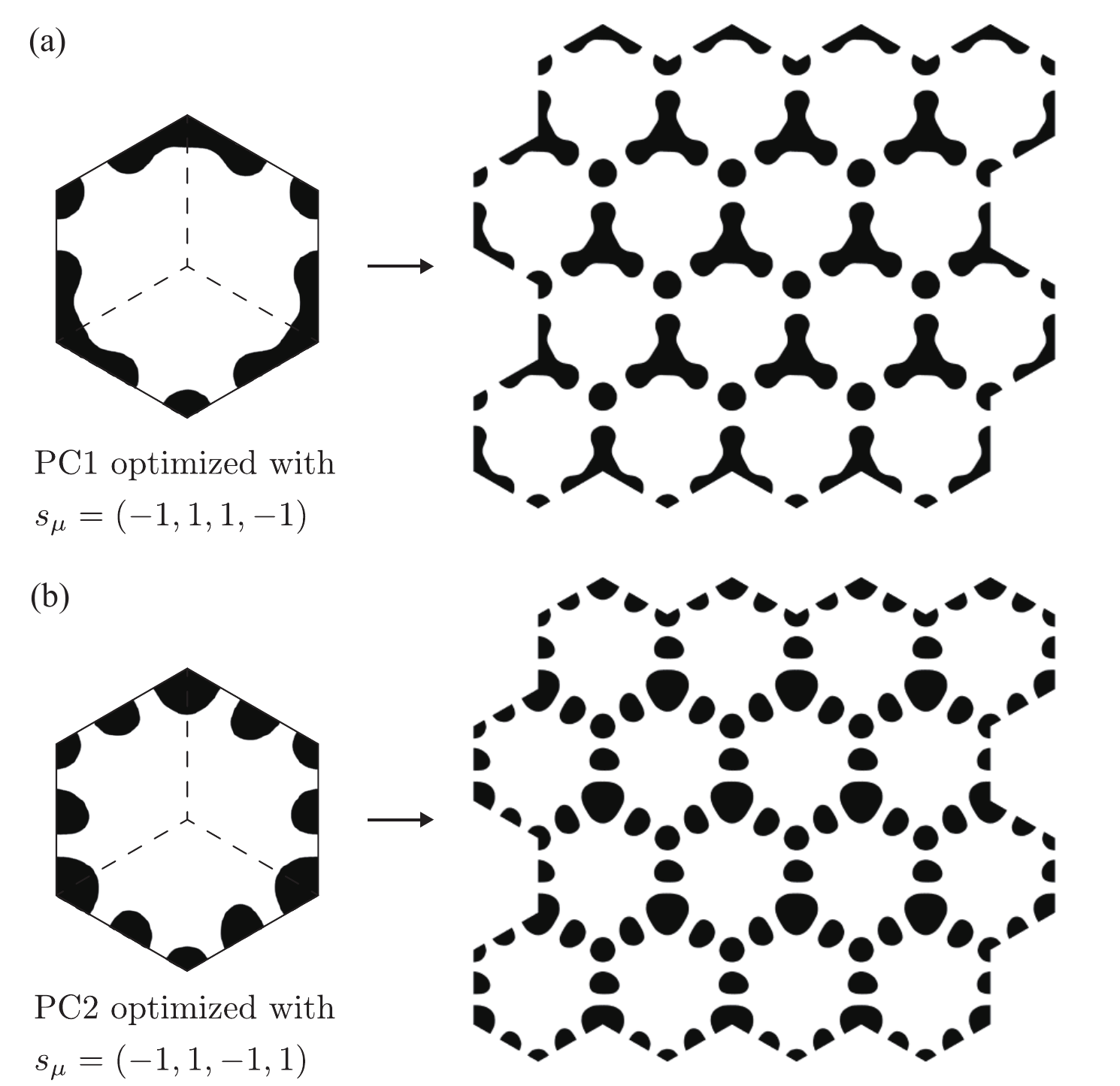}
    \caption{Optimized unit cells and photonic crystals: (a) PC1 optimized with pseudospin $s_\mu = (-1, 1, 1, -1)$; (b) PC2 optimized with pseudospin $s_\mu = (-1, 1, -1, 1)$.}
    \label{fig:result}
\end{figure}

\begin{figure*}[t]
    \centering
    \includegraphics[width=0.85\textwidth]{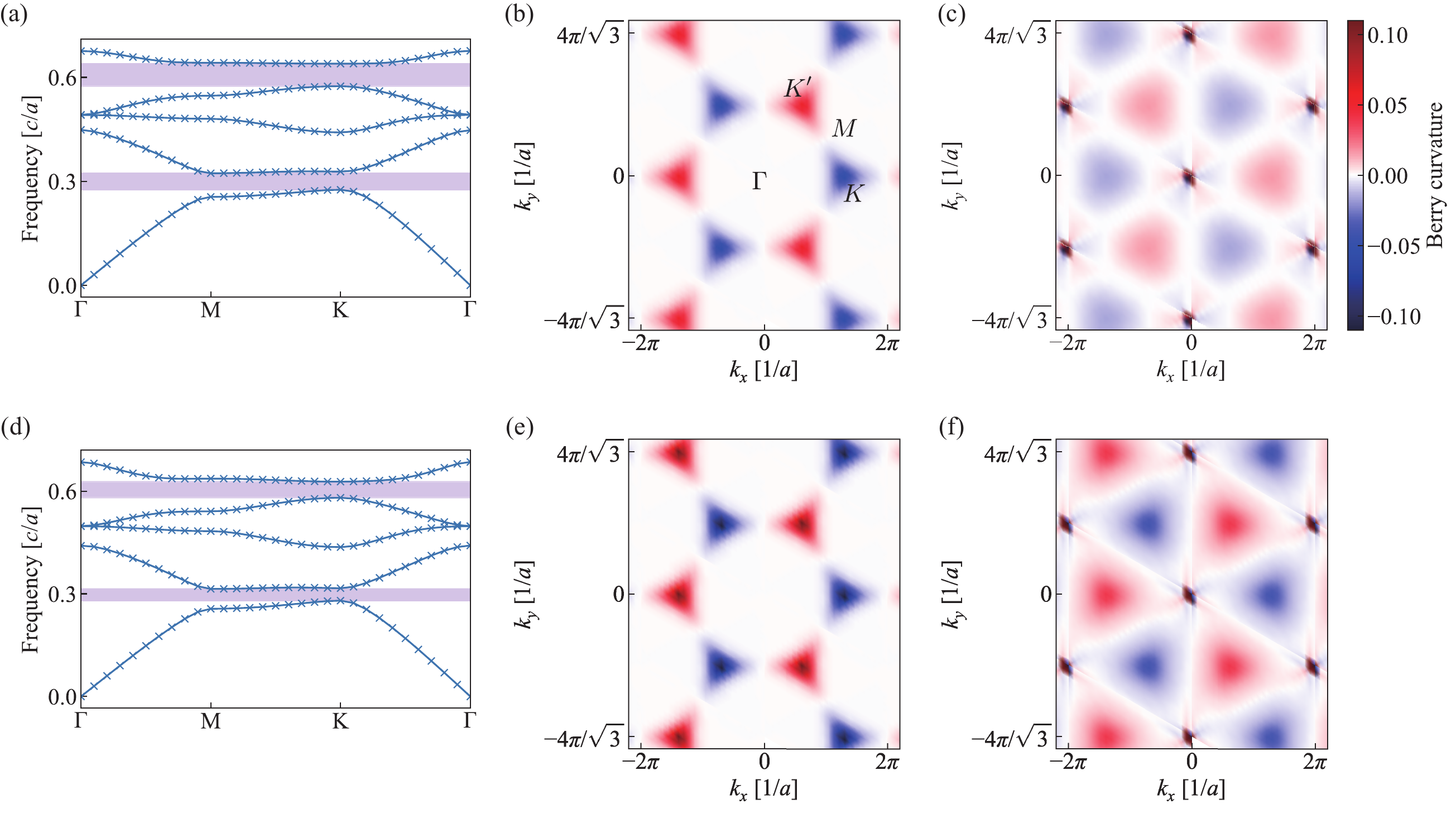}
    \caption{Band structures (a, d) and Berry curvatures (b, c, e, f). (a) and (d): Band structures of PC1 and PC2, respectively. Band gaps are shown in purple. (b) and (e): Berry curvatures of the first bands of PC1 and PC2, respectively. (c) and (f): Berry curvatures of the fourth bands of PC1 and PC2, respectively.}
    \label{fig:band}
\end{figure*}

\begin{figure*}[t]
    \centering
    \includegraphics[width=0.65\textwidth]{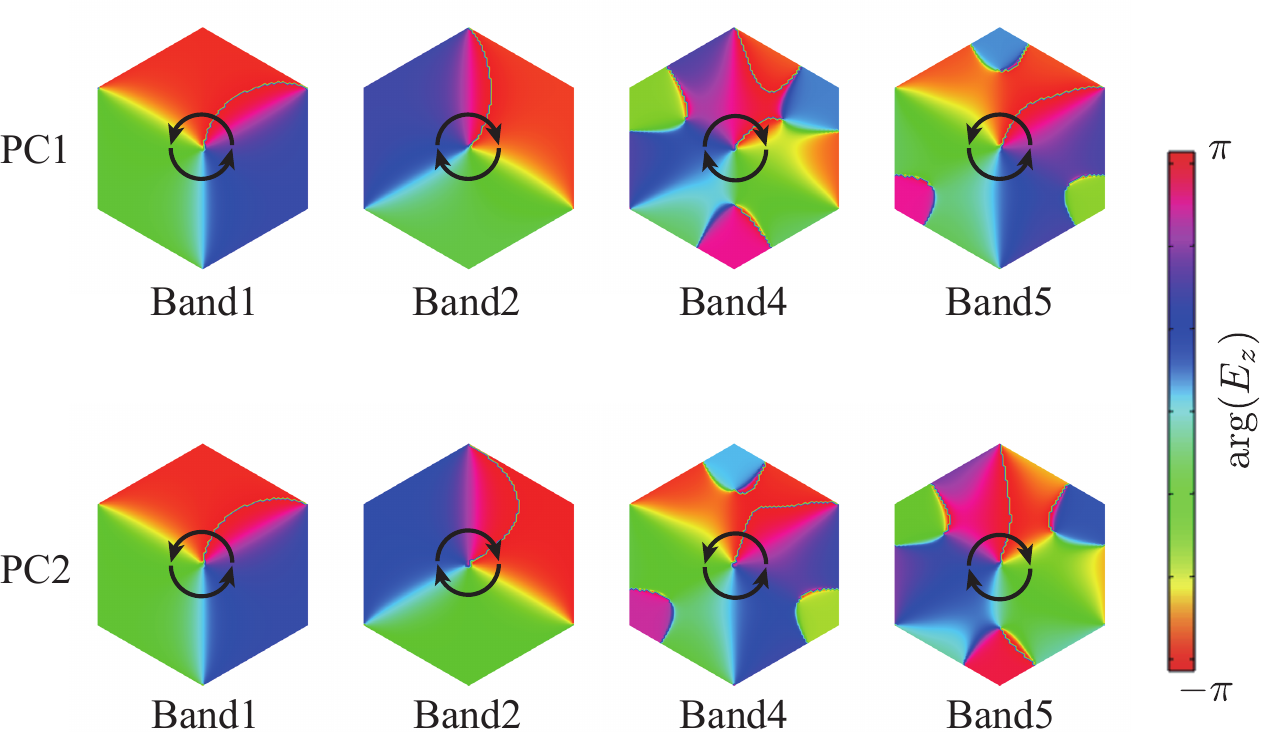}
    \caption{Phases of eigenmodes of the first, second, fourth, and fifth bands (denoted as Band1, Band2, Band4, Band5, respectively) at the K point for PC1 and PC2.}
    \label{fig:phase}
\end{figure*}

\section{Results and discussion} \label{sec:result}
\vspace*{-4pt}

\subsection{Optimized unit cell structures} \label{sec:result_unitcell}
\vspace*{-4pt}
In this study, two PCs (PC1 and PC2) were designed with different target pseudospin settings.
Specifically, $s_\mu = (-1, 1, 1, -1)$ for PC1 and $s_\mu = (-1, 1, -1, 1)$ for PC2 were used.
This means that the same target pseudospin value was set for the bands forming the first band gap and the opposite target pseudospin for those forming the second band gap.
As discussed later, this setting of target pseudospin values enables the design of a frequency-dependent waveguide router.
Figure~\ref{fig:result} illustrates the optimized PCs.
Optimized structures were successfully obtained depending on the target pseudospins.
The spatial inversion symmetry is broken in both structures; this is consistent with a common mechanism for opening the Dirac cone, leading to a valley Hall effect~\cite{iwamoto2021recent}.

To assess the validity of the optimized PCs, we computed the band structures and Berry curvatures of the optimized PCs using eigenvalue analysis over the first Brillouin zone.
We evaluated the Berry curvatures using discrete Wilson loops based on the method in Ref.~\cite{blanco2020tutorial}.
Specifically, we discretized the first Brillouin zone into a $24 \times 24$ regular grid in the momentum space and calculated the Berry phase for each grid cell, as follows.
\begin{align}
    \phi(\boldsymbol{k}_{\mu, \nu}) &= -\mathrm{Im} \log \left( \langle \bar{E}_z^\ast(\boldsymbol{x}; \boldsymbol{k}_{\mu, \nu}) | \varepsilon_\mathrm{r} \bar{E}_z(\boldsymbol{x}; \boldsymbol{k}_{\mu,(\nu+1)}) \rangle \right. \nonumber \\
    &\quad \left. \times \langle \bar{E}_z^\ast(\boldsymbol{x}; \boldsymbol{k}_{\mu,(\nu+1)}) | \varepsilon_\mathrm{r} \bar{E}_z(\boldsymbol{x}; \boldsymbol{k}_{(\mu+1),(\nu+1)}) \rangle \right. \nonumber \\
    &\quad \left. \times \langle \bar{E}_z^\ast(\boldsymbol{x}; \boldsymbol{k}_{(\mu+1),(\nu+1)}) | \varepsilon_\mathrm{r} \bar{E}_z(\boldsymbol{x}; \boldsymbol{k}_{(\mu+1),\nu}) \rangle \right. \nonumber \\
    &\quad \left. \times \langle \bar{E}_z^\ast(\boldsymbol{x}; \boldsymbol{k}_{(\mu+1),\nu}) | \varepsilon_\mathrm{r} \bar{E}_z(\boldsymbol{x}; \boldsymbol{k}_{\mu,\nu}) \rangle \right),
\end{align}
where $k_{\mu, \nu}$ is the wave vector at the $(\mu, \nu)$ grid point and $\bar{E}_z$ is the eigenstate of the electric field.
This equation can be used to find the Berry curvatures because the Berry phase calculated using the Wilson loop on a small grid cell is proportional to the Berry curvature at that cell~\cite{blanco2020tutorial}.

Figure~\ref{fig:band} shows the band structures and Berry curvatures of the optimized PCs.
Both PC1 and 2 exhibit two band gaps around the target frequencies, as shown in Figures~\ref{fig:band}(a) and (d): the first and second bands form the first band gap, whereas the fourth and fifth bands form the second band gap.
As shown in Figures~\ref{fig:band}(b), (c), (e), and (f), PC1 and PC2 have nonzero Berry curvatures around the K and K' points.
Specifically, PC1 and PC2 have the same (opposite) sign of the Berry curvature at the first (fourth) band around the K and K' points, which means that they have the same valley topological phase in the first band, but the distinct valley topological phases in the fourth band.
Note that the curvatures behave singularly around the $\Gamma$ point for the fourth band because the third and fourth bands are degenerate at the $\Gamma$ point as shown in Figures~\ref{fig:band} (a) and (d).
The signs of the Berry curvatures at the K point correspond to the signs of the target pseudospins, demonstrating the validity of the proposed method.

Figure~\ref{fig:phase} illustrates the phases of eigenmode of the first, second, fourth, and fifth bands at the K point for PC1 and PC2.
Both PCs have the same rotational direction for the first and second bands; they have the opposite rotational directions for the fourth and fifth bands, which is consistent with the setting of the target pseudospins.
This suggests that our formulation of the TSAM using a magnetic field effectively generates vorticity in the phase distribution of the electric field, i.e., pseudospin states.
In summary, our proposed method successfully obtained two PCs that are spontaneously equipped with the target topological properties through optimization in the frequency domain.

\subsection{Valley kink states} \label{sec:result_supercell}
\vspace*{-4pt}

\begin{figure*}[t]
    \centering
    \includegraphics[width=\textwidth]{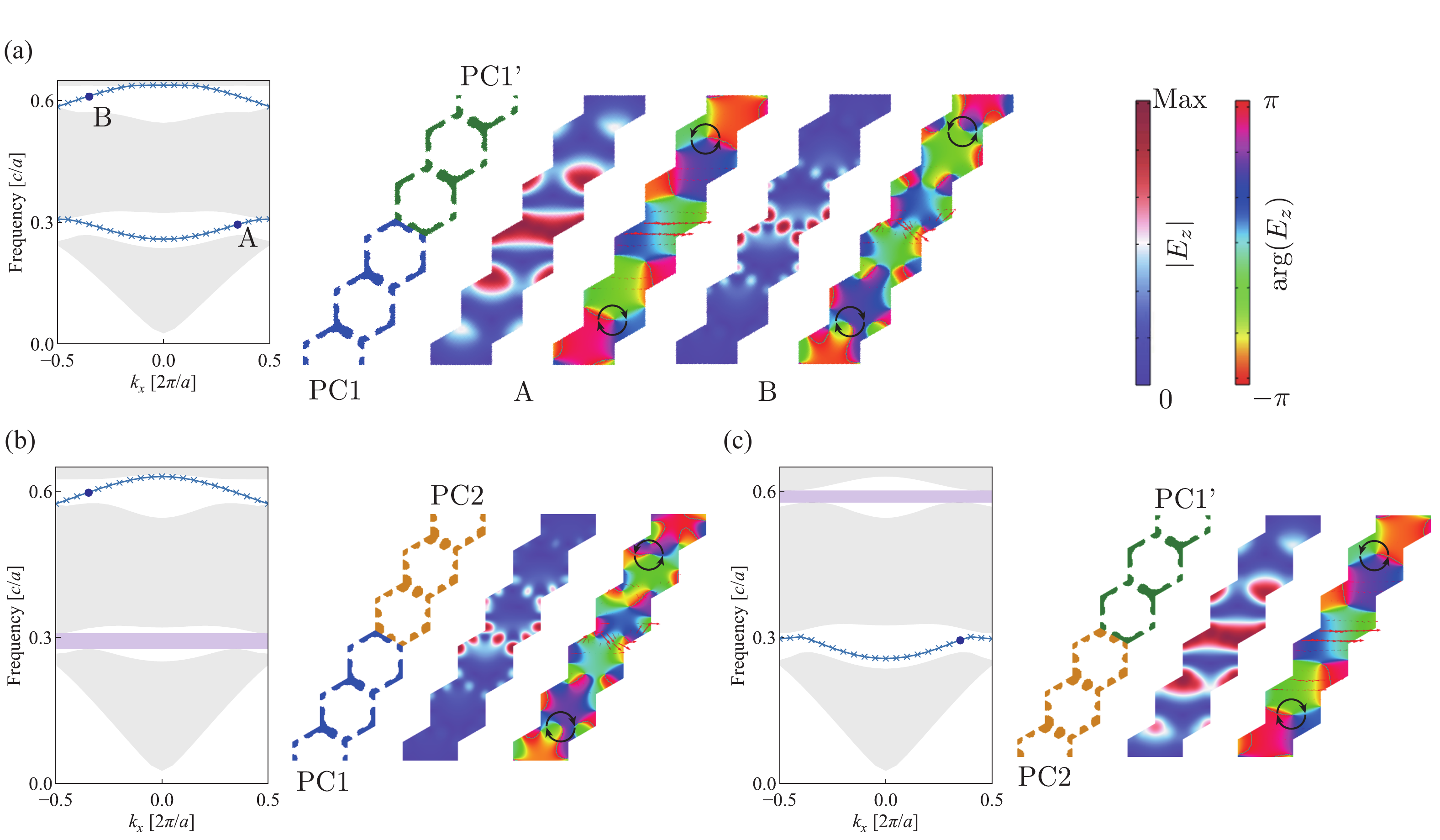}
    \caption{Band structures of the supercells along with their eigenstates and corresponding phase distribution. (a) Supercells composed of PC1 (blue) and PC1' (green) cells. (b) Supercells comprising PC1 (blue) and PC2 (orange) cells. (c) Supercells comprising PC2 (orange) and PC1' (green) cells. 
    In the band structure, purple regions are band gaps, blue lines indicate bands exhibiting edge states, and blue points correspond to each eigenstate and its phase.
    The red arrows in the phase distribution represent Poynting vectors.
    The black arrows represent rotational direction of the electric-field phase distribution.}
    \label{fig:supercell}
\end{figure*}

Next, we investigated valley kink states in the boundary of PC1 and PC2. 
Based on the results in Section~\ref{sec:result_unitcell}, PC1 and PC2 have the same valley topological phase in the first band but distinct valley topological phases in the fourth band.
This implies that the interface between PC1 and PC2 has valley kink states in the higher band gap formed between the fourth and fifth bands of these PCs (as we later confirm).
To form valley kink states in the lower band gap, we prepared a spatially inverted version of PC1 (hereinafter called PC1').
Because PC1 and PC1' are related by spacial inversion, PC1' has distinct valley topological phases in both the first and fourth bands (See Supplementary Information for the band structure, Berry curvatures and phase distributions of eigenmodes of PC1').
Consequently, by appropriately combining two PCs from the set (PC1, PC1', PC2), valley kink states can be generated at their interfaces within their lower and/or higher band gap of these PCs.

To confirm that valley kink states in the band gap of the PCs, we computed the band structures of three supercells: PC1 and PC1', PC1 and PC2, and PC2 and PC1'.
Each supercell comprises six unit cells (three of one type, and three of the other; see Figure~\ref{fig:supercell}).
The band structures were computed by eigenvalue analysis under the Floquet periodic boundary condition.
Figure~\ref{fig:supercell} illustrates the band structures of the supercells along with their eigenstates and corresponding phases.
The blue lines in the band structures represent bands exhibiting valley kink states, whereas the purple regions indicate band gaps.
Each eigenstate and its phase correspond to a blue point in the band structure.
The red arrows in the phase distribution represent Poynting vectors.
As shown in Figure~\ref{fig:supercell}(a), the interface between PC1 and PC1' has valley kink states within both the lower and the higher band gap of these PCs;
these states have the same rotational direction of the electric-field phase distribution (black arrows) and the same direction of energy transport (red arrows).
Figure~\ref{fig:supercell}(b) shows that the interface between PC1 and PC2 has valley kink states only within their higher band gap.
The valley kink state has the same pseudospin and the same energy-transport direction as does the supercell composed of PC1 and PC1' (Figure~\ref{fig:supercell}(a)).
Furthermore, the eigenstate and phase distribution of the supercell composed of PC1 and PC2 (Figure~\ref{fig:supercell}(b)) are almost identical to those at point B in the higher band of the supercell composed of PC1 and PC1' (Figure~\ref{fig:supercell}(a)).
Figure~\ref{fig:supercell}(c) shows that the interface between PC2 and PC1' has valley kink states only within their lower band gap.
The valley kink state has the same pseudospin and the same energy-transport direction as does the supercell composed of PC1 and PC1' (Figure~\ref{fig:supercell}(a)).
Furthermore, the eigenstate and phase distribution of the supercell composed of PC2 and PC1' (Figure~\ref{fig:supercell}(c)) are almost identical to those at point A in the lower band of the supercell composed of PC1 and PC1' (Figure~\ref{fig:supercell}(a)).
Thus, it was confirmed that valley kink states can be generated within either the lower band gap (PC2, PC1'), the higher band gap (PC1, PC2), or both band gaps (PC1, PC1') by a pairwise combination of PC1, PC1', and PC2.

\subsection{Frequency-dependent waveguide router}
\vspace*{-4pt}
\begin{figure*}[t]
    \centering
    \includegraphics[width=0.95\textwidth]{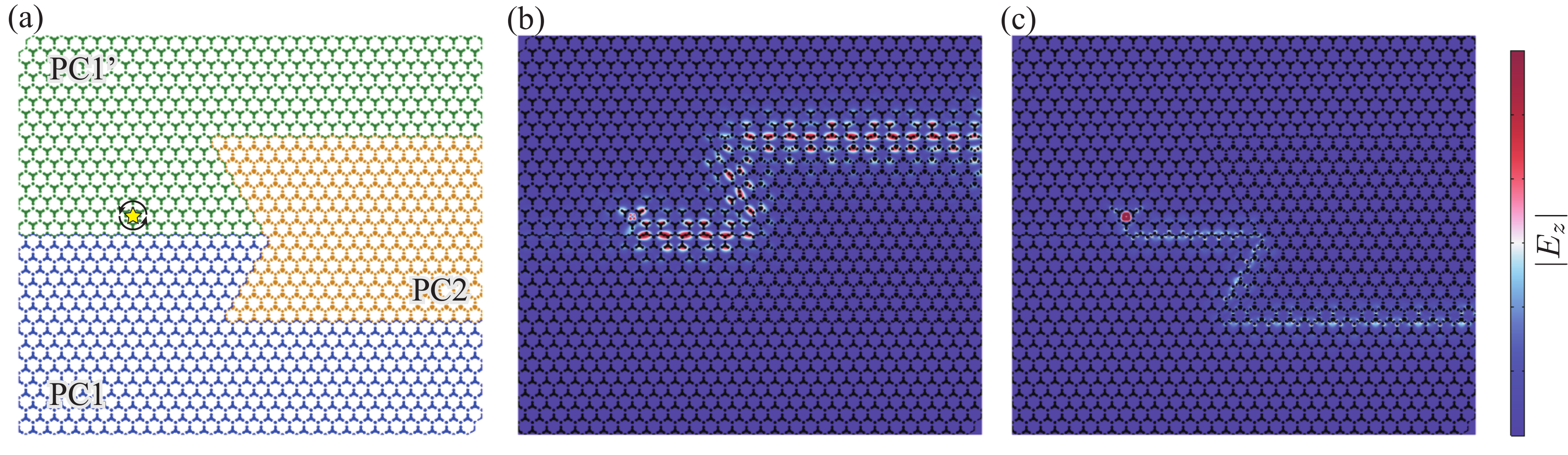}
    \caption{Valley photonic crystal waveguide (a) and its frequency-dependent light paths (b, c). (a) Waveguide structure includes three types of photonic crystal: PC1 (blue region), spatially inverted PC1 (PC1'; green region), and PC2 (orange region). Light is provided at the yellow star. (b) Magnitude of electric fields at frequency $0.295~[c/a]$. (c) Magnitude of electric fields at frequency $0.6~[c/a]$. The color range is the same in (b) and (c); however, it has been adjusted to emphasize the typical electric field intensities throughout the simulation area, as the field is highly concentrated near the input port.}
    \label{fig:waveguide}
\end{figure*}

\begin{figure}[t]
    \centering
    \includegraphics[width=\columnwidth]{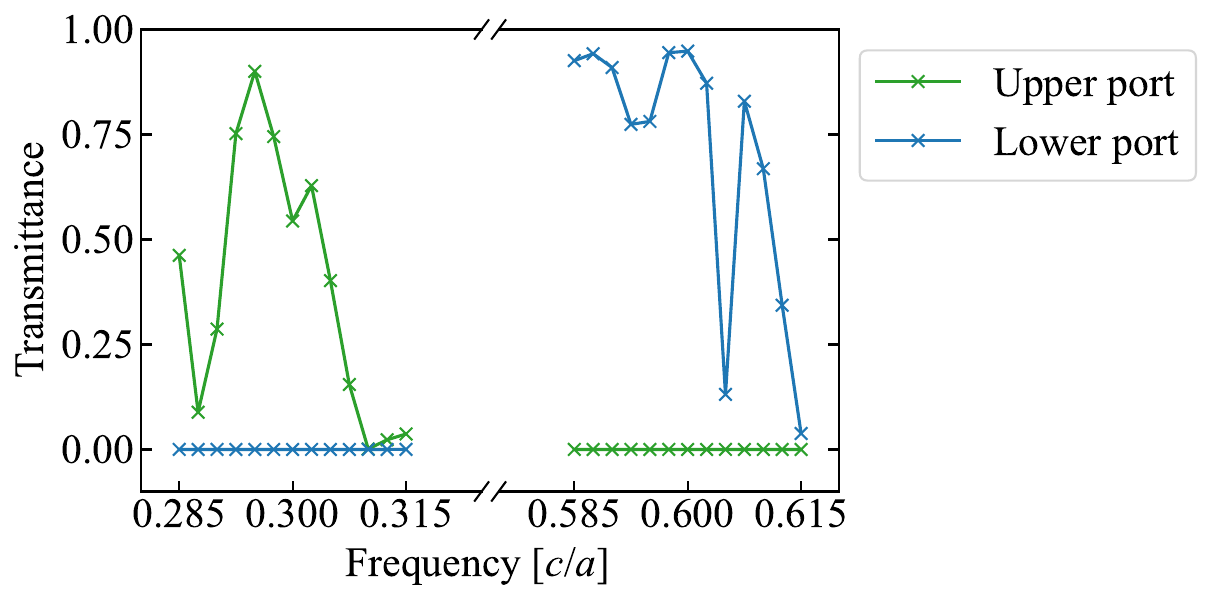}
    \caption{Transmittance of the two ports of the valley photonic crystal waveguide.}
    \label{fig:transmittance}
\end{figure}

To demonstrate the utility and applicability of the proposed method, we present a VPC waveguide that enables light to propagate along frequency-dependent paths~\cite{tang2020frequency,wei2021frequency}.
Based on the results in Section~\ref{sec:result_supercell}, when the interface between PC1 and PC2 is connected to that between PC1 and PC1', light is expected to propagate in the higher-frequency band via the valley kink state, whereas in the lower-frequency band the presence of a band gap prevents light from propagating into the interface between PC1 and PC2.
Similarly, when the interface between PC2 and PC1' is connected to that between PC1 and PC1', light can propagate only in the lower-frequency band because of the presence of the band gap.

On the basis of the above discussion, we designed the VPC waveguide shown in Figure~\ref{fig:waveguide}(a).
It consisted of three types of photonic crystals, PC1 (blue region), PC1' (green region), and PC2 (orange region).
The PCs were placed considering their pseudospins and formed a sharp bend to demonstrate robust light propagation.
The waveguide had four-line sources in a square array (at the yellow star in the figure); these applied electric currents with anticlockwise phase differences to excite the valley kink state, as shown in Figure~\ref{fig:supercell}.
A perfectly matched layer was set around the PCs to mitigate reflection from the boundaries of the computational domain.

Figures~\ref{fig:waveguide}(b) and (c) illustrate the magnitude of the electric field at frequencies of $0.295~[c/a]$ and $0.6~[c/a]$, respectively.
These figures clearly show that the input light propagates unidirectionally along the frequency-dependent paths (the upper path for the frequency $0.295~[c/a]$ and the lower path for the frequency $0.6~[c/a]$) in a robust manner, even with sharp bends.
To evaluate the transmittance, the output energy was computed as the boundary integral of the $x$ component (horizontal component) of the Poynting vector, i.e., $P_\mathrm{out, i} := \int_{\Gamma_\mathrm{out, i}} \vec{S} \cdot \vec{e}_x \mathrm{d}\Gamma$ for $i=1,2$.
Here, $\Gamma_\mathrm{out, i}$ is the output port of the waveguide, $\vec{S}$ is the Poynting vector, and $\vec{e}_x$ is the basis vector along with the $x$ (horizontal) axis.
The input energy was computed as the power emitted by the electric current, i.e., $P_\mathrm{in} := - \int_{\Omega_\mathrm{in}} \mathrm{Re} (J_z^\ast E_z) \mathrm{d}\Omega$ where $\Omega_\mathrm{in}$ is the area in which the electric current was applied and $J_z$ is the applied current.
The transmittance was $P_\mathrm{out, 1} / P_\mathrm{in} = 0.90$ at frequency $0.295~[c/a]$ and $P_\mathrm{out, 2} / P_\mathrm{in} = 0.95$ at the frequency $0.6~[c/a]$.
The high transmittance for both paths indicates that the waveguide composed of the optimized PCs can function as a frequency-dependent router in a highly robust and efficient manner.

To examine this in more detail, we evaluated the transmittance spectra in Figure~\ref{fig:waveguide}(a) for frequencies of around $0.3~[c/a]$ and $0.6~[c/a]$, as is shown in Figure~\ref{fig:transmittance}.
The upper port, i.e., the waveguide composed of PC2 and PC1', has a relatively high transmittance around the frequency of $0.3~[c/a]$ whereas the lower port, i.e., the waveguide composed of PC1 and PC2, exhibit a relatively high transmittance around the frequency of $0.6~[c/a]$.
This is consistent with the band structures of the supercells, discussed in Section~\ref{sec:result_supercell}. 
In other words, light propagates along the interface between PC1 and PC2 in the higher-frequency band while along the interface between PC1' and PC2 in the lower-frequency band.
To achieve a broadband waveguide, it would be necessary to tailor the topological bands and band gaps of the supercells; we plan to address this problem in future work.

\section{Conclusions} \label{sec:conclusions}
\vspace*{-4pt}
In this paper, we propose inverse designs for VPCs in which the valley topological phases of PCs are generated spontaneously.
Our formulation considers arbitrary pseudospin states and is based on frequency-domain analysis, which makes the optimization problem computationally efficient.
We successfully obtain two PC designs with two band gaps around the target frequencies.
They have the target valley topological phases: the same valley topological phases in the first band and distinct valley topological phases in the fourth band.
Through band structure analysis for supercells consisting of optimized PCs, we have confirmed that by appropriately combining optimized PCs, valley kink states can exist within their lower and/or higher band gap.
To demonstrate the proposed method, we have designed a waveguide consisting of two designed PCs.
Numerical experiments demonstrate that light propagates unidirectionally along frequency-dependent paths with high transmittance under the excitation of chiral current sources.
The proposed method represents a new way of designing VPCs and will widen their applicability to various functional devices.

\section*{Data availability statement}
The datasets generated during and/or analyzed during the current study are available from the corresponding author on reasonable request.

\section*{Funding information}
None declared.

\section*{Conflict of interest}
Authors state no conflict of interest.

\bibliography{reference}

\clearpage
\pagebreak
\onecolumngrid
\begin{center}
\textbf{\large Supplementary information: Inverse design of dual-band valley-Hall topological photonic crystals with arbitrary pseudospin state}
\end{center}
\setcounter{section}{0}
\setcounter{equation}{0}
\setcounter{figure}{0}
\setcounter{table}{0}
\setcounter{page}{1}
\makeatletter
\renewcommand{\thesection}{S\arabic{section}}
\renewcommand{\theequation}{S\arabic{equation}}
\renewcommand{\thefigure}{S\arabic{figure}}
\renewcommand{\bibnumfmt}[1]{[S#1]}
\renewcommand{\citenumfont}[1]{S#1}

\section{Relationship between pseudospin and transverse spin angular momentum}

Let $\phi_x = \arg (H_x)$ and $\phi_y = \arg (H_y)$.
Substituting $H_x = \lvert H_x \rvert e^{i \phi_x}$ and $H_y = \lvert H_y \rvert e^{i \phi_y}$ into the transverse spin angular momentum defined in the main text yields
\begin{align}
    S&:= -\frac{2}{\lvert \Omega_\mathrm{s} \rvert} \mathrm{Im} \left( \int_{\Omega_\mathrm{s}} \frac{H_x^\dagger H_y}{ H_x^\dagger H_x + H_y^\dagger H_y} \mathrm{d} \Omega \right) \nonumber \\
    &= -\frac{1}{\lvert \Omega_\mathrm{s} \rvert} \mathrm{Im} \left( \int_{\Omega_\mathrm{s}} \frac{2\lvert H_x \rvert \lvert H_y \rvert e^{i (\phi_y - \phi_x)} }{ \lvert H_x \rvert^2 + \lvert H_y \rvert^2} \mathrm{d} \Omega \right) \nonumber \\
    &= -\frac{1}{\lvert \Omega_\mathrm{s} \rvert} \int_{\Omega_\mathrm{s}} \frac{2\lvert H_x \rvert \lvert H_y \rvert \sin (\phi_y - \phi_x) }{ \lvert H_x \rvert^2 + \lvert H_y \rvert^2} \mathrm{d} \Omega.
\end{align}
From the arithmetic mean-geometric mean inequality, we have
\begin{align}
    &\frac{ \lvert H_x \rvert^2 + \lvert H_y \rvert^2}{2} \geq \lvert H_x \rvert \lvert H_y \rvert,
\end{align}
with the equality condition $\lvert H_x \rvert = \lvert H_y \rvert$.
Consequently, we obtain the following inequality:
\begin{align}
    |S| &= \frac{1}{\lvert \Omega_\mathrm{s} \rvert} \int_{\Omega_\mathrm{s}} \frac{2\lvert H_x \rvert \lvert H_y \rvert \sin (\phi_y - \phi_x) }{ \lvert H_x \rvert^2 + \lvert H_y \rvert^2} \mathrm{d} \Omega \nonumber \\
    &\leq \frac{1}{\lvert \Omega_\mathrm{s} \rvert} \int_{\Omega_\mathrm{s}} \frac{2\lvert H_x \rvert \lvert H_y \rvert }{ \lvert H_x \rvert^2 + \lvert H_y \rvert^2} \mathrm{d} \Omega \nonumber \\
    &\leq 1,
\end{align}
with the equality condition $\lvert H_x \rvert = \lvert H_y \rvert$ and $\lvert \phi_y - \phi_x \rvert = \pi/2$, which corresponds to the maximal transverse spin angular momentum associated with a circularly polarized magnetic field.
The sign of $S$ represents that of transverse spin angular momentum.

Now, we relate the maximal transverse spin angular momentum to the pseudospin of the electric field, i.e., the vorticity of the phase of the electric field.
First, we show that there exists the vorticity of the phase of the out-of-plane electric field when the in-plane magnetic field is circularly polarized under a reasonable assumption.
From the relationship $(H_x, H_y) = (i \omega^{-1} \partial E_z / \partial y, -i \omega^{-1} \partial E_z / \partial x)$, the equality condition of $|S|=1$ implies $\lvert \partial E_z / \partial y \rvert = \lvert \partial E_z / \partial x \rvert$ and $\arg (\partial E_z / \partial x) - \arg (\partial E_z / \partial y) = \pm \pi/2$.
By coordinate transformation, $|S|=1$ also implies $\lvert r^{-1} \partial E_z / \partial \theta \rvert = \lvert \partial E_z / \partial r \rvert$ and $\arg (\partial E_z / \partial r) - \arg (r^{-1} \partial E_z / \partial \theta) = \pm \pi/2$ where $(r, \theta)$ is the cylindrical coordinate system.
Let $E_z = |E_z|e^{i \phi}$ where $\phi$ is the phase of the electric field.
The spatial derivatives in the cylindrical coordinate are given as
\begin{align}
    \frac{\partial E_z}{\partial r} &= \left( \frac{\partial |E_z|}{\partial r} + i \frac{\partial \phi}{\partial r} |E_z| \right) e^{i \phi} \\
     \frac{\partial E_z}{\partial \theta} &= \left( \frac{\partial |E_z|}{\partial \theta} + i \frac{\partial \phi}{\partial \theta} |E_z| \right) e^{i \phi}.
\end{align}
Their arguments satisfy
\begin{align}
    \pm \frac{\pi}{2} &= \arg \left( \frac{\partial E_z}{\partial r} \right) - \arg \left( \frac{1}{r} \frac{\partial E_z}{\partial \theta} \right) \nonumber \\
    &= \arg \left( \frac{\partial |E_z|}{\partial r} + i \frac{\partial \phi}{\partial r} |E_z| \right) - \arg \left( \frac{1}{r}\frac{\partial |E_z|}{\partial \theta} + i \frac{1}{r}\frac{\partial \phi}{\partial \theta} |E_z| \right) \nonumber \\
    &= \arg \left( \left( \frac{1}{r} \frac{\partial |E_z|}{\partial r} \frac{\partial |E_z|}{\partial \theta} + |E_z|^2 \frac{1}{r} \frac{\partial \phi}{\partial r} \frac{\partial \phi}{\partial \theta} \right)
 + i |E_z| \frac{1}{r} \left( \frac{\partial \phi}{\partial r}
 \frac{\partial |E_z|}{\partial \theta} - \frac{\partial \phi}{\partial \theta} 
 \frac{\partial |E_z|}{\partial r} \right) \right) \nonumber \\
    &= \arg \left( \left( \frac{1}{r} \frac{\partial |E_z|}{\partial r} \frac{\partial |E_z|}{\partial \theta} + |E_z|^2 \frac{1}{r} \frac{\partial \phi}{\partial r} \frac{\partial \phi}{\partial \theta} \right)
 + i |E_z|^2 \frac{1}{r} \left( \frac{\partial \phi}{\partial r}
 \frac{\partial \log |E_z|}{\partial \theta} - \frac{\partial \phi}{\partial \theta} 
 \frac{\partial \log |E_z|}{\partial r} \right) \right). \label{eq:argument}
\end{align}
For this equation to be satisfied, the real part of the complex variable on the right-hand side must be zero, which reads
\begin{align}
     &\frac{1}{r}\frac{\partial |E_z|}{\partial r} \frac{\partial |E_z|}{\partial \theta} + |E_z|^2 \frac{1}{r} \frac{\partial \phi}{\partial r} \frac{\partial \phi}{\partial \theta} = 0
\end{align}
Rearranging this equation, we obtain
\begin{align}
     \frac{1}{r} \frac{\partial \log |E_z|}{\partial r} \frac{\partial \log |E_z|}{\partial \theta} + \frac{1}{r} \frac{\partial \phi}{\partial r} \frac{\partial \phi}{\partial \theta} = 0. \label{eq:argument_condition}
\end{align}
From the condition of the amplitude, the following also holds:
\begin{align}
    &\left( \frac{\partial |E_z|}{\partial r} \right)^2 + \left( \frac{\partial \phi}{\partial r} |E_z| \right)^2 = \left( \frac{1}{r} \frac{\partial |E_z|}{\partial \theta} \right)^2 + \left( \frac{1}{r} \frac{\partial \phi}{\partial \theta} |E_z| \right)^2,
\end{align}
which is reformulated as
\begin{align}
    \left( \frac{\partial \log|E_z|}{\partial r} \right)^2 + \left( \frac{\partial \phi}{\partial r} \right)^2 = \left( \frac{1}{r} \frac{\partial \log|E_z|}{\partial \theta} \right)^2 + \left( \frac{1}{r} \frac{\partial \phi}{\partial \theta} \right)^2. \label{eq:amplitude_condition}
\end{align}
Combining the conditions for the argument and amplitude in Equations~\eqref{eq:argument_condition} and \eqref{eq:amplitude_condition}, we obtain the following two conditions, either of which must be satisfied:
\begin{align}
    &\frac{1}{r} \frac{\partial \log |E_z|}{\partial \theta} = \frac{\partial \phi}{\partial r}, ~ -\frac{\partial \log |E_z|}{\partial r} = \frac{1}{r} \frac{\partial \phi}{\partial \theta} \label{eq:condition1} \\
    &- \frac{1}{r} \frac{\partial \log |E_z|}{\partial \theta} = \frac{\partial \phi}{\partial r}, ~ \frac{\partial \log |E_z|}{\partial r} = \frac{1}{r} \frac{\partial \phi}{\partial \theta}. \label{eq:condition2}
\end{align}
Now, we consider the sign of the argument difference, i.e., the sign of $S$.
The imaginary part of the right-hand side in Equation~\eqref{eq:argument} corresponds to the sign as follows.
\begin{align}
    \begin{cases}
        \frac{\pi}{2} = \arg \left( \frac{\partial E_z}{\partial r} \right) - \arg \left( \frac{1}{r} \frac{\partial E_z}{\partial \theta} \right) & \text{ if } \frac{\partial \phi}{\partial r}
 \frac{\partial \log |E_z|}{\partial \theta} - \frac{\partial \phi}{\partial \theta} 
 \frac{\partial \log |E_z|}{\partial r} > 0 \\
        -\frac{\pi}{2} = \arg \left( \frac{\partial E_z}{\partial r} \right) - \arg \left( \frac{1}{r}\frac{\partial E_z}{\partial \theta} \right) & \text{ if } \frac{\partial \phi}{\partial r}
 \frac{\partial \log |E_z|}{\partial \theta} - \frac{\partial \phi}{\partial \theta} 
 \frac{\partial \log |E_z|}{\partial r} < 0.
    \end{cases} \label{eq:argument_sign}
\end{align}
Substituting the condition in Equation~\eqref{eq:condition1} into the condition in Equation~\eqref{eq:argument_sign}, we obtain 
\begin{align}
    \frac{\partial \phi}{\partial r}
 \frac{\partial \log |E_z|}{\partial \theta} - \frac{\partial \phi}{\partial \theta} 
 \frac{\partial \log |E_z|}{\partial r} =
 \frac{1}{r} \left( \frac{\partial \log |E_z|}{\partial \theta} \right)^2 + r \left(
 \frac{\partial \log |E_z|}{\partial r} \right)^2 > 0,
\end{align}
which means that Equation~\eqref{eq:condition1} corresponds to the condition $\frac{\pi}{2} = \arg (\partial E_z / \partial r) - \arg (r^{-1} \partial E_z / \partial \theta)$.
Similarly, from Equations~\eqref{eq:condition2} and \eqref{eq:argument_sign}, we obtain 
\begin{align}
    \frac{\partial \phi}{\partial r}
 \frac{\partial \log |E_z|}{\partial \theta} - \frac{\partial \phi}{\partial \theta} 
 \frac{\partial \log |E_z|}{\partial r} =
 - \frac{1}{r} \left( \frac{\partial \log |E_z|}{\partial \theta} \right)^2 - r \left(
 \frac{\partial \log |E_z|}{\partial r} \right)^2 < 0,
\end{align}
which means that Equation~\eqref{eq:condition2} corresponds to the condition $-\frac{\pi}{2} = \arg (\partial E_z / \partial r) - \arg (r^{-1} \partial E_z / \partial \theta)$.
Assuming that $\partial \log |E_z| / \partial r > 0$, Equation~\eqref{eq:condition1} implies $\partial \phi / \partial \theta = - r\partial \log |E_z|/ \partial r < 0$, while Equation~\eqref{eq:condition2} implies $\partial \phi / \partial \theta = r\partial \log |E_z|/ \partial r > 0$.
The monotonic decrease $\partial \phi / \partial \theta < 0$ (increase $\partial \phi / \partial \theta > 0$) of the phase of $E_z$ along with the azimuth means the vorticity of the phase of the electric field, that is the pseudospin of the electric field.
Therefore, we can relate the transverse spin angular momentum to the pseudospin of the electric field as follows.
\begin{align}
    \frac{\partial \phi}{\partial \theta} < 0 & \text{ if } S = 1 \\
    \frac{\partial \phi}{\partial \theta} > 0 & \text{ if } S = -1,
\end{align}
under the assumption of $\partial \log |E_z| / \partial r > 0$.
This assumption is valid in the vicinity of the origin because materials composing photonic crystals are located on perimeters of unit cells, on which the electric fields are concentrated.

Next, we show that an in-plane magnetic field has maximal transverse spin angular momentum, i.e., circular polarization when there exists the vorticity of the phase of the out-of-plane electric field under a reasonable assumption.
Let $E_z$ be assumed, for example, that $E_z(r, \theta) = E(r) e^{i l \theta}$ where $(r, \theta)$ is the cylindrical coordinate.
The spatial derivatives in the cylindrical coordinate are given as
\begin{align}
    \frac{\partial E_z}{\partial r} &= E'(r)e^{i l\theta} \\
    \frac{\partial E_z}{\partial \theta} &= il E(r) e^{i l\theta}. 
\end{align}
That is, the magnetic fields in the cylindrical coordinate are represented as
\begin{align}
    H_r &= i \omega^{-1} r^{-1} \frac{\partial E_z}{\partial \theta} = -\omega^{-1} r^{-1} lE(r)e^{i l\theta} \nonumber \\
    H_\theta &= -i \omega^{-1} \frac{\partial E_z}{\partial r} =-i \omega^{-1} E'(r)e^{i l\theta}.
\end{align}
Therefore, we obtain
\begin{align}
    \phi_r &= \frac{1 + \mathrm{sgn}(l)}{2} \pi + l \theta \\
    \phi_\theta &= -\frac{\pi}{2} + l \theta,
\end{align}
which yields
\begin{align}
    \phi_r - \phi_\theta = \mathrm{sgn}(l) \pi / 2.
\end{align}
That is, the in-plane magnetic field is elliptically polarized.
Furthermore, let us assume $E(r) = \alpha r + O(r^2)$ since we evaluate the $S(\omega)$ in the vicinity of the origin $r=0$, then we obtain 
\begin{align}
    H_r &= - \omega^{-1} l \alpha e^{i l \theta} + O(r^2) \\
    H_\theta &= -i \omega^{-1} \alpha e^{il \theta} + O(r^2),
\end{align}
which means that $|H_r| = |H_\theta| + O(r^2)$ when $l = \pm1$. 
This is exactly the condition that the magnetic field is circularly polarized.

\section{Topological properties of spatially inverted version of PC1}
\begin{figure}[H]
    \centering
    \includegraphics[width=0.85\textwidth]{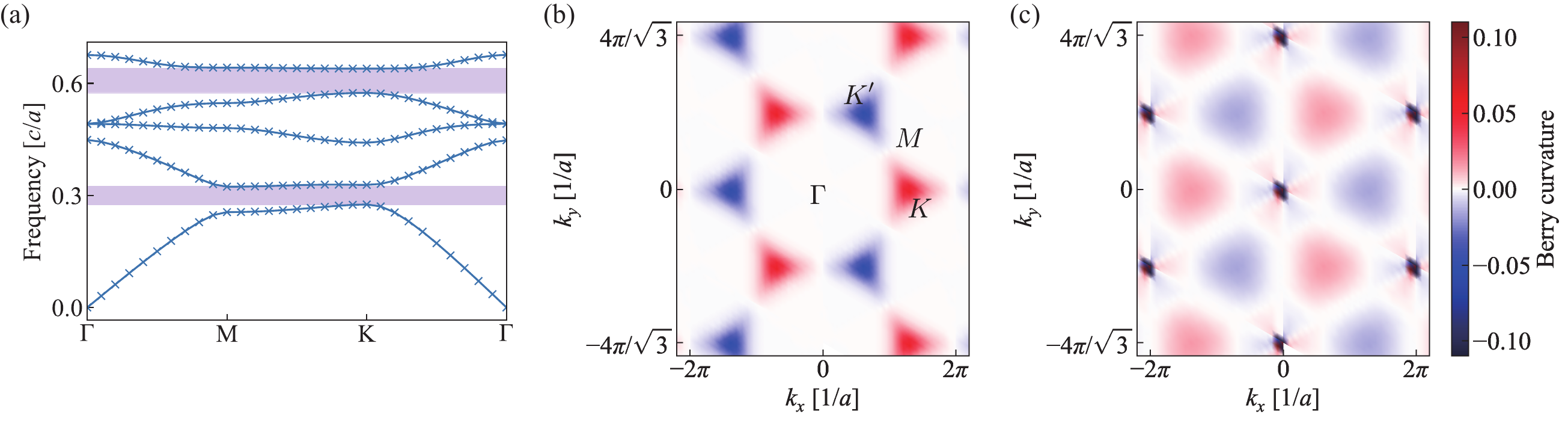}
    \caption{Band structures (a) and Berry curvatures (b, c). (a): Band structure of PC1' (spatially inverted version of PC1). Band gaps are shown in purple. (b) and (c): Berry curvatures for the first and fourth bands, respectively.}
    \label{fig:band_inverted}
\end{figure}

Figure~\ref{fig:band_inverted} shows the band structure and the Berry curvatures of PC1' (i.e., the spatially inverted version of PC1).
PC1' has the same band gap as PC1 while PC1 and PC1' have the opposite sign of the Berry curvatures around K and K' points for both the first and fourth bands.
This means that PC1 and PC1' have distinct valley topological phases both in the first and fourth bands.

\begin{figure}[H]
    \centering
    \includegraphics[width=0.65\textwidth]{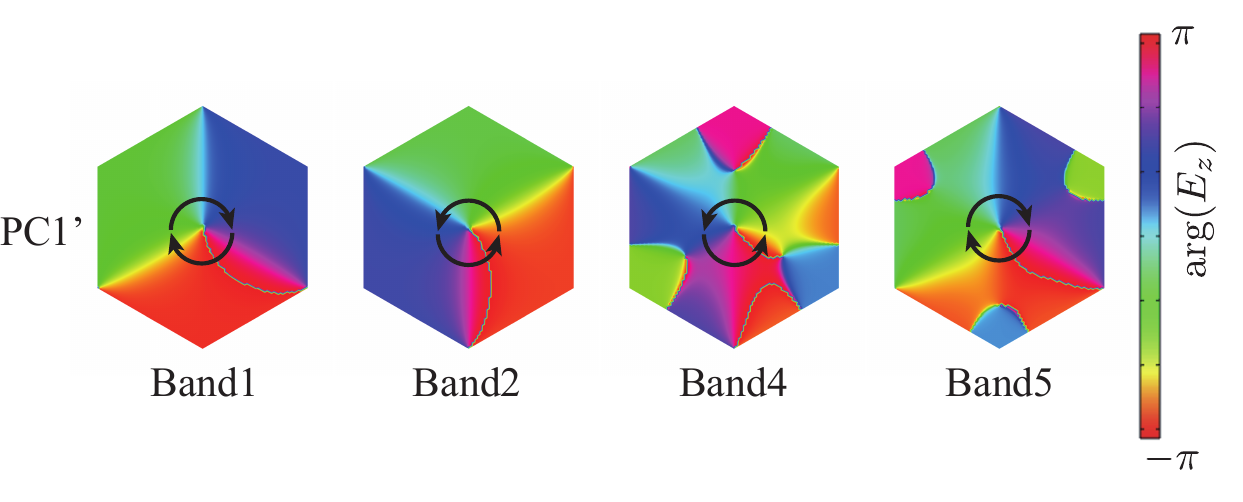}
    \caption{Phases of eigenmodes of first, second, fourth, and fifth bands (denoted by Band1, Band2, Band4, Band5, respectively) at the K point for PC1' (spatially inverted version of PC1).}
    \label{fig:phase_inverted}
\end{figure}

Figure~\ref{fig:phase_inverted} illustrates the phase distributions of eigenmodes of the first, second, fourth, and fifth bands at the K point for PC1'.
PC1 and PC1' have the opposite rotational direction for all bands, which enables the design of a frequency-dependent waveguide router using PC1, PC1', and PC2.

\end{document}